# $g_{\phi\pi\gamma}$ coupling constant in light cone QCD sum rules

C.Aydin[*], A.H.Yilmaz[†], and M.Bayar

*Physics Department, Karadeniz Technical University, 61080 Trabzon, Turkey*
[*]coskun@ktu.edu.tr
[†]hakany@ktu.edu.tr

The coupling constant of $\phi \to \pi\gamma$ decay is calculated using light cone QCD sum rules. A comparison of our result with the ones existing in the literature is presented.

Radiative transitions between pseudoscalar (P) and vector (V) mesons have been an important subject in low-energy hadron physics for more than three decades. These transitions have been considered as phenomenological quark models, potential models, bag models, and effective Lagrangian methods[1,2]. Since they determine the strength of the hadron interactions, among the characteristics of the strong interaction processes $g_{\phi\pi\gamma}$ coupling constant plays one of the most important roles,. In the quark models, V → P+γ decays (V = φ, ρ, ω)  ; P = π, η, η′) are reduced by the quark magnetic moment with transition s = 1 → s = 0, where s is the total spin of the $q\bar{q}$ − system (in the corresponding meson). These quantities can  be calculated directly from QCD. Low-energy hadron interactions are governed by nonperturbative QCD so that it is very difficult to get the numerical values of the coupling constants from the first principles. For that reason a semiphenomenological method of QCD sum rules can be used, which nowadays is the standart tool for studying of various characteristics of hadron interactions. On the other hand, vector meson-pseudoscalar meson-photon VPγ–vertex  also plays a role in photoproduction reactions of vector mesons on nucleons. It should be notable that in these decays (V→Pγ) the four-momentum of the pseudoscalar meson P is time-like, $P'^2 > 0$, whereas in the pseudoscalar exchange amplitude contributing to the photoproduction of vector mesons it is space-like $P'^2 < 0$. Therefore, it is of interest to study the effective coupling constant $g_{VP\gamma}$ from another point of view as well. The coupling constant of $\phi \to \pi\gamma$  decay was calculated[3] in the QCD sum rules and obtained as $g_{\phi\pi\gamma} = 0.125$.

The QCD sum rule[4-7] is a framework widely used to investigate hadronic properties in terms of QCD degrees of freedom. In this method, it is crucial to represent a correlation function through a dispersion relation. This is because QCD calculations of the correlation function through the operator product expansion (OPE) can be done only in deep space-like regions, whereas the hadronic parameters are defined by the nonanalytic structure existing in time-like regions. Through a dispersion relation, the calculated correlation function can be matched with the hadronic parameters. At present, there are two methods to construct QCD sum rules when an external field is present: the conventional approach, relying on the short-distance expansion, and the light-cone QCD sum rule (LCQSR), based on the expansion along the light-cone. The method of QCD sum rules represent one of few methods capable of making predictions in the low- and medium-energy hadron physics starting basically from the QCD Lagrangian. The only phenomenological input parameters are the values of two or three quark and gluon condensates characterizing the properties of the physical vacuum. The field of application of the sum rules has been extended remarkably during the past twenty years[5]. Application of this method to polarization operators gives a determination of masses and couplings of low lying mesonic[4,6] and baryonic[4] states.

The light cone QCD sum rules are based on the operator product expansion on the light cone, which is an expansion over the twist of the operators rather than dimensions as in the traditional QCD sum rules. In this expansion, the main contribution comes from the lowest twist operators. The matrix elements of the nonlocal operators between a hadronic state and the vacuum define the hadronic wave functions which are the principle nonperturbative inputs into the sum rules. The applications of the light cone QCD sum rules to study hadronic properties can be found in[8-11] and references therein. In this paper we apply the light cone QCD sum rules to determine the $g_{\phi\pi\gamma}$ coupling constant.

In order to determine the coupling constant $g_{\phi\pi\gamma}$ in the framework of the light cone QCD sum rules, we consider the two point correlation function:

$$\Pi_\mu(p_1,p_2) = i\int d^4x\, e^{ip.x} <\gamma(q)|T\{J_\mu^\phi(0)J_5^{\pi^0}(x)\}|0> \qquad (1)$$

We choose the interpolating current for the $\phi$ and $\pi$ mesons as $J_\mu^\phi = \sin\theta j_\mu^{\omega_0} = \frac{1}{6}(\bar{u}\gamma_\mu u + \bar{d}\gamma_\mu d)\sin\theta$, and $J_5^{\pi^0} = \frac{1}{2}[\bar{u}(i\gamma_5)u - \bar{d}(i\gamma_5)d]$ respectively. The mixing angle that used is determined by Bramon et al. as $((\theta = 3.4\pm 0.2)^\circ$ [12]. The physical part of the sum rules can be attained by inserting a complete set of one-meson states into the equation (1):

$$\Pi_{\mu\nu} = \sum \frac{\langle 0|J_5^\pi(x)|\pi(p_2)\rangle}{p_2^2 - m_\pi^2}\langle \pi(p_2)|\phi(p_1)\rangle_\gamma \frac{\langle \phi(p_1)|J_\mu^\phi(0)|0\rangle}{p_1^2 - m_\phi^2} \qquad (2)$$

where $p_1 = p_2 + q$ and $q$ is the photon momentum. The matrix element entering Eq.(2) is defined as

$$\langle 0|J_\mu^\phi|\phi\rangle = m_\phi f_\phi^{(u)} \varepsilon_\mu^\phi \qquad (3)$$

$$\langle 0|J_5|\pi\rangle = \lambda_\pi \qquad (4)$$

where $\varepsilon^\phi$ is the $\phi$ meson polarization vector. In SU(3) symmetry limit $f_\phi^{(s)} = 6 f_\phi^{(u)}$ where upper s and u mean quark content of $\phi$ meson. Note that $f_\phi^{(s)}$ is found from $\phi \to e^+e^-$ decay (see below). The coupling constant $g_{\phi\pi\gamma}$ for $\phi \to \pi\gamma$ decay is defined as follows

$$\langle \pi(p_2)|\phi(p_1)\rangle_\gamma = e\varepsilon_{\mu\nu\alpha\beta}\varepsilon_\mu^\phi p_{2\nu}\varepsilon_\alpha^\gamma q_\beta F(q^2) \qquad (5)$$

where $\varepsilon^\lambda$ is the photon polarization vector and $q$ is the photon momentum. Since the photon is real, we need the value of $F(q^2)$ only at the point $q^2 = 0$.

We can use an alternative parametrization for the $\phi\pi\gamma$ vertex, i.e.,

$$L_{int} = -\frac{e}{m_\phi} g_{\phi\pi\gamma} \varepsilon_{\mu\nu\alpha\beta} (\partial_\mu \phi_\nu - \partial_\nu \phi_\mu)(\partial_\alpha A_\beta - \partial_\beta A_\alpha)\pi^0. \tag{6}$$

Comparing Eqns. (5) and (6) we see that

$$F(q^2 = 0) \equiv \frac{g_{\phi\pi\gamma}}{m_\phi}. \tag{7}$$

Having used Eqs. (2) - (6), for the physical part of the sum rules we get

$$\Pi^{ph}_{\mu\nu} = g_{\phi\pi\gamma} \frac{\lambda_\pi}{p_2^2 - m_\pi^2} \frac{f_\phi^{(u)} \varepsilon_\alpha^\gamma}{p_1^2 - m_\phi^2} \varepsilon_{\mu\nu\alpha\beta} p_{2\nu} q_\beta. \tag{8}$$

In this calculation the full light quark propagator with both perturbative and nonperturbative contribution is used, and it is given as[13]

$$iS(x,0) = \langle 0|T\{\bar{q}(x)q(0)\}|0\rangle$$

$$= i\frac{\not{x}}{2\pi^2 x^4} - \frac{\langle\bar{q}q\rangle}{12} - \frac{x^2}{192} m_0^2 \langle\bar{q}q\rangle - ig_s \frac{1}{16\pi^2} \int_0^1 du \left\{ \frac{\not{x}}{x^2} \sigma_{\mu\nu} G^{\mu\nu}(ux) - 4iu \frac{x_\mu}{x^2} G^{\mu\nu}(ux)\gamma_\nu \right\} + ... \tag{9}$$

where terms proportional to light quark mass $m_u$ or $m_d$ are neglected. After straightforwrad computation we have

$$T_\mu(p,q) = \frac{1}{6}\int d^4 x e^{ipx} \left\{ -B\langle\gamma(q)|\bar{u}(x)\gamma_\mu\gamma_5 u(0)|0\rangle - \frac{1}{2} A[\varepsilon_{\alpha\mu\sigma\phi} x_\alpha]\langle\gamma(q)|\bar{u}(x)\sigma_{\phi\sigma} u(0)|0\rangle \right\} \sin\theta \tag{10}$$

where $A = \frac{i}{2\pi^2 x^4}$ and $B = -\frac{1}{12}\langle\bar{u}u\rangle - \frac{m_0^2}{192}\langle\bar{u}u\rangle x^2$. In calculation of the nonperturbative contributions by the OPE on the light cone one needs to know the matrix elements of nonlocal operators between vacuum and the photon states: i.e. $\langle\gamma(q)|\bar{q}\Gamma_i q|0\rangle$ where $\Gamma_i$ is an arbitrary Dirac matrix. These matrix elements can be expand in terms of photon wave functions with definite twist. In calculations we neglect twist 3 three particle photon wave function. Twist two and twist four photon wave function are defined as[14]

$$\langle\gamma(q)|\bar{q}\gamma_\alpha\gamma_5 q|0\rangle = \frac{f}{4} e_q e \varepsilon_{\alpha\beta\phi\sigma} \varepsilon^\beta q^\phi x^\sigma \int_0^1 du e^{iuqx} \Psi(u),$$

$$\langle\gamma(q)|\bar{q}\sigma_{\alpha\beta}q|0\rangle = ie_q <\bar{q}q> \int_0^1 du\, e^{iuqx}$$

$$\times\{(\varepsilon_\alpha q_\beta - \varepsilon_\beta q_\alpha)[\chi\varphi(u) + x^2[g_1(u) - g_2(u)]] + [qx(\varepsilon_\alpha x_\beta - \varepsilon_\beta x_\alpha) + \varepsilon x(x_\alpha q_\beta - x_\beta q_\alpha)]g_2(u)\} \quad (11)$$

In these equations $e_q$ is the corresponding quark charge. $\chi$ is the magnetic susceptibility, $\Psi(u)$ and $\varphi(u)$ are leading twist two and $g_1(u)$ and $g_2(u)$ are the twist four photon wave functions.

After evaluating the Fourier transform for the M1 structure and then performing the double Borel transformation with respect to the variables $Q_1^2 = -p_1^2$ and $Q_2^2 = -p_2^2$, we finally obtain the following sum rule for the coupling constant $g_{\phi\pi\gamma}$

$$g_{\phi\pi\gamma} = \frac{m_\phi(e_u - e_d)<\bar{u}u>}{\lambda_\pi\lambda_\phi} e^{m_\phi^2/M_1^2} e^{m_\pi^2/M_2^2} \times \{-\chi\varphi(u_0)M^2 E_0(s_0/M^2) + 2(g_1(u_0) - g_2(u_0))\}\sin\theta, \quad (12)$$

where the function

$$E_0(s_0/M^2) = 1 - e^{-s_0/M^2} \quad (13)$$

is the factor used to subtract the continuum, $s_0$ being the continuum threshold, and

$$u_0 = \frac{M_1^2}{M_1^2 + M_2^2}, \quad M^2 = \frac{M_1^2 M_2^2}{M_1^2 + M_2^2} \quad (14)$$

with $M_1^2$ and $M_2^2$ are the Borel parameters in the $\phi$ and $\pi$ channels, $\lambda_\phi = m_\phi f_\phi$.

For the numerical evaluation of the sum rule we used the values $<\bar{u}u> = -0.014\,\text{GeV}^3$ [5], $m_\phi = 1.02\,\text{GeV}$, $m_\pi = 0.138\,\text{GeV}$ [5], and for the magnetic susceptibility $\chi = -3.4\,\text{GeV}^{-2}$ [16]. Since $\lambda_\pi = f_\pi \frac{m_\pi^2}{m_u + m_d}$, then $\lambda_\pi$ is given as $\lambda_\pi = 0.18\,\text{GeV}^2$. We note that neglecting the electron mass the $e^+e^-$ decay width of $\phi$ meson is given as $\Gamma(\phi \to e^+e^-) = \frac{4\pi\alpha^2}{3}\frac{\lambda_\phi^2}{m_\phi^3}$. Then using the value from the experimental leptonic decay width of $\phi$ [4], we obtain the value $\lambda_\phi = (0.079 \pm 0.001)\,\text{GeV}^2$ for the overlap amplitude $\phi$ meson. For the photon wave functions[17, 18] we make use of the following expressions:

$$\varphi(u_0) = 6u_0(1 - u_0),$$

$$g_1(u_0) = -\frac{1}{8}(1 - u_0)(3 - u_0), \quad (15)$$

$$g_2(u_0) = -\frac{1}{4}(1-u_0)^2$$

In Fig. 1 we show the dependence of the coupling constant $g_{\phi\pi\gamma}$ on the Borel parameter $M_1^2$ at some different values of the continuum threshold as $s_0 = 1.8$, 2.0, and 2.2 GeV$^2$. Since the Borel mass $M_2^2$ is an auxiliary parameter and the physical quantities should not depend on it, one must look for the region where $g_{\phi\pi\gamma}$ is practically independent of $M_1^2$. We determined that this condition is satisfied in the interval $2.0\ GeV^2 \leq M_1^2 \leq 2.6\ GeV^2$. The variation of the coupling constant $g_{\phi\pi\gamma}$ as a function of Borel parameters for different values $M_1^2$ and $M_2^2$ at $s_0 = 2.0$ GeV$^2$ are shown in Fig. 2. Examination of this figure points out that the sum rule is rather stable with these reasonable variations of $M_1^2$ and $M_2^2$. We choose the middle value $M_2^2 = 2.2$ GeV$^2$ for the Borel parameter in its interval of variation and obtain the coupling constant $g_{\phi\pi\gamma}$ as $g_{\phi\pi\gamma} = 0.13$.

We would like to compare our predictions on $g_{\phi\pi\gamma}$ with experimental result. The decay width of $\phi \to \pi^0\gamma$ decay is given by

$$\Gamma(\phi \to \pi^0\gamma) = \frac{\alpha}{24}\frac{(m_\phi^2 - m_\pi^2)^3}{m_\phi^5}g_{\phi\pi\gamma}^2. \tag{16}$$

Using the experimental value of $\Gamma(\phi \to \pi^0\gamma)$ [19], the $g_{\phi\pi\gamma}$ coupling constant is obtained from above equation as $g_{\phi\pi\gamma} = 0.135 \pm 0.002$ which is very close to the sum rule estimation.

**Acknowledgments**

The authors would like to thank Profs.T.M.Aliev and O.Yilmaz for a very useful discussion and for careful examination of the manuscript. This work was partly supported by the Research Fund of Karadeniz Technical University, under grand contract no 2002.111.001.2.

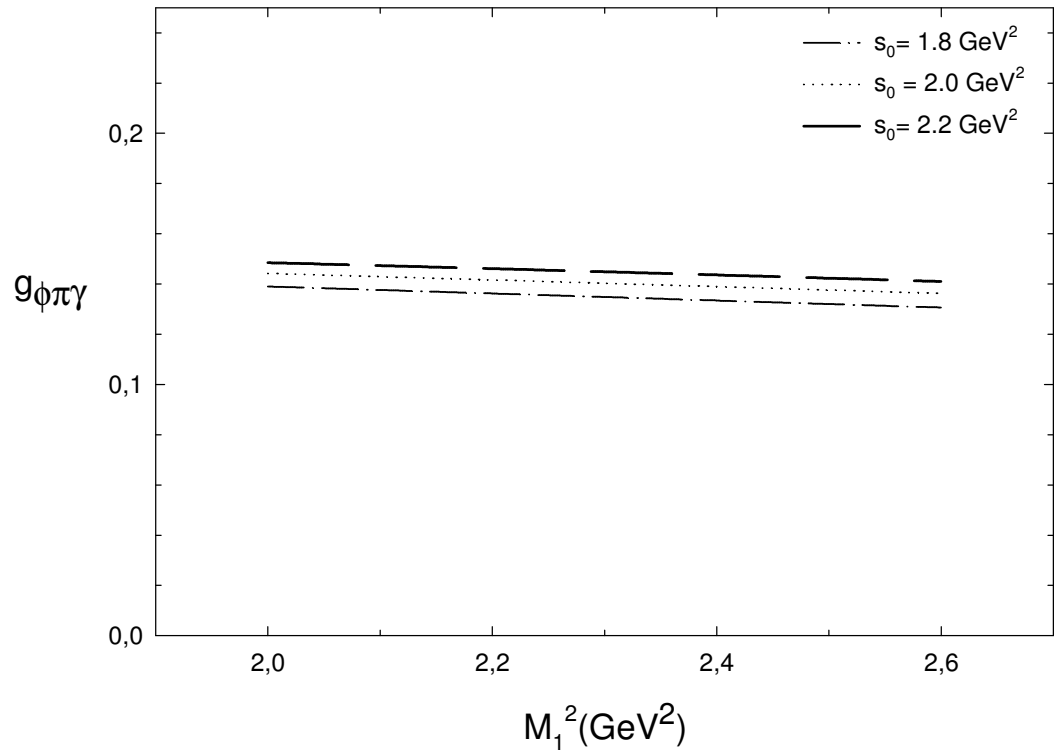

Fig. 1. The coupling constant $g_{\phi\pi\gamma}$ as a function of the Borel parameter $M_1^2$ for different values of the threshold parameters $s_0$ with $M_2^2=2.0$ GeV$^2$.

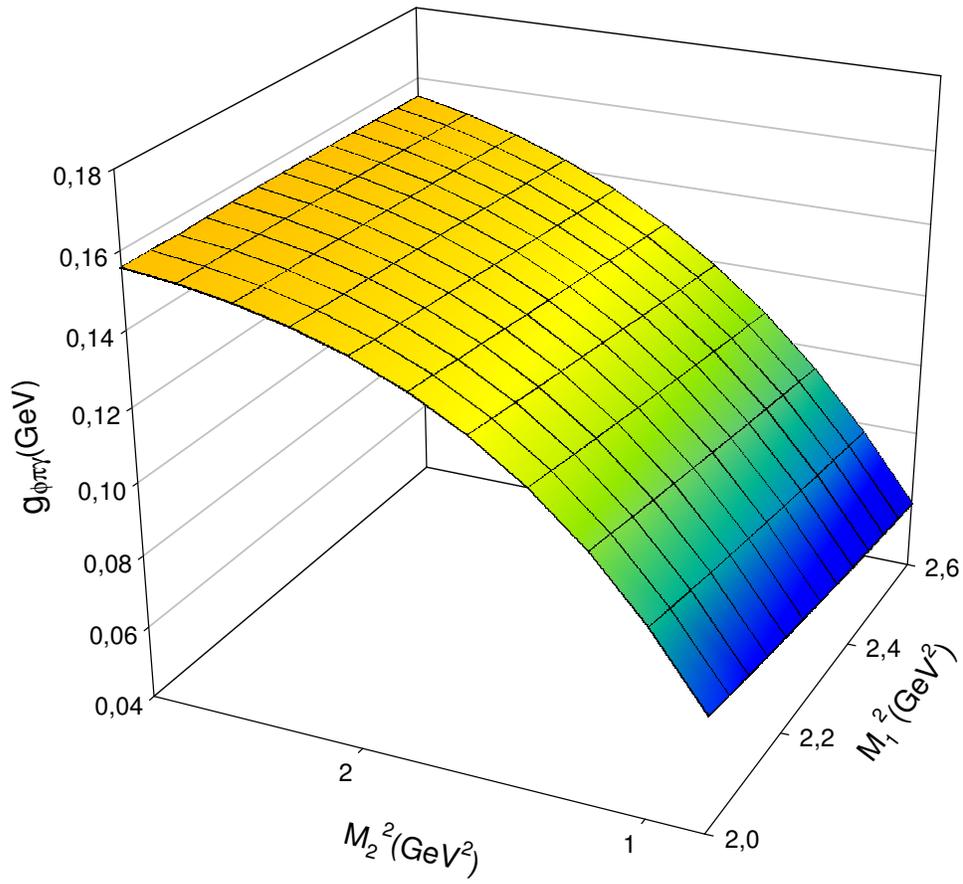

Fig. 2. Coupling constant $g_{\phi\pi\gamma}$ as a function of the Borel parameters for different values $M_1^2$ and $M_2^2$ at $s_0$=2.0 GeV$^2$.